\documentclass[10pt,final,technote,letter,oneside,twocolumn]{IEEEtran}

\usepackage[noadjust]{cite}
\usepackage{srcltx}
\usepackage{psfrag}
\usepackage{epsfig}
\usepackage{epstopdf}
\usepackage[tbtags,sumlimits,nointlimits,reqno]{amsmath}
\usepackage{amssymb}
\usepackage{xcolor}
\usepackage{float}
\usepackage{subfigure}
\usepackage{soul}
\usepackage{footnote}
\usepackage{amsmath}
\graphicspath{{Figs/}}

\begin{document}

\title{Enhancement of Time-Reversal Subwavelength Wireless Transmission Using Pulse Shaping}

\author{Shuai Ding, Shulabh Gupta, Rui Zang, Lianfeng Zou, Bing-Zhong Wang,~\IEEEmembership{Member,~IEEE}\\
        Christophe Caloz,~\IEEEmembership{Fellow,~IEEE}
\thanks{S. Ding, Shulabh Gupta, L. Zou and C. Caloz are with the Department
of Electrical Engineering and Poly-Grames research center, \'{E}cole Polytechnique de Montr\'{e}al, Montr\'{e}al, Qu\'{e}bec, H3T 1J4, Canada. e-mail: shuai.ding@polymtl.ca; christophe.caloz@polymtl.ca.}
\thanks{R. Zang and B.-Z. Wang are with the Institute of Applied Physics, University
of Electronic Science and Technology of China, Chengdu 610054, China. e-mail: bzwang@uestc.edu.cn}}
\markboth{}%
{Shell \MakeLowercase{\textit{et al.}}: Bare Demo of IEEEtran.cls for Journals}

\maketitle

\begin{abstract}
A novel time-reversal subwavelength transmission technique, based on pulse shaping circuits (PSCs), is proposed. This technique removes the need for complex or electrically large electromagnetic structures by generating channel diversity via pulse shaping instead of angular spectrum transformation. It is shown that, compared to our previous time-reversal system based on chirped delay lines, the PSC approach offers greater flexibility and larger possible numbers of channels, i.e. ultimately higher transmission throughput. The PSC based time-reversal system is also demonstrated experimentally.

\end{abstract}

\IEEEpeerreviewmaketitle
\section{Introduction}\label{Sec:Introduction}

\IEEEPARstart{T}{i}me reversal is an adaptive electromagnetic transmission technique with applications in many areas, such as for instance wireless communication, imaging and sensing~\cite{Bogert-Timereversal-1957,Fink-Basic-1992,Nguyen-2005-Communlett-potential,Gu-2014-Timereversalantenna}. A typical time-reversal system operates as follows. First, channel sounding signals are emitted from multiple sources. Next, the channel sounding signals are received by a time reversal mirror (TRM) and time reversed. Finally, the time-reversed signals are re-transmitted by the TRM. If the channel from the TRM to the sources is reciprocal, the time-reversed signals retrace the incoming path and focus to the location of the initial source.

Recent research has shown that time-reversal can lead to electromagnetic far-field sub-wavelength focusing, which represents a novel technique to overcome the Rayleigh-diffraction limit~\cite{Fink-2007-Science-Farfield}. This unique mechanism has led to interesting applications in wireless sub-wavelength transmission and far-field super-resolution detection~\cite{Ge-Super-2009,Liao-subwavelength-2009,Fink-2013-plr}. One approach to achieve time-reversal sub-wavelength focusing is to transform evanescent waves into propagating waves via appropriate manipulation of the angular spectrum of the involved electromagnetic waves. This has been done by placing randomly distributed metal scatterers in the near field of the sources~\cite{Fink-2007-Science-Farfield,Ge-Super-2009,Fusco-conjugation-2010,Fink-2002-Timereversalplr,Gu-2013-Timereversalantenna}, or electromagnetically large multilayered dielectric structures in front of the TRM~\cite{Liao-subwavelength-2009}. However, the approach is hardly reproducible~\cite{Ge-Super-2009} while the latter requires very bulky elements. Another approach to achieve time-reversal sub-wavelength focusing is based on Radio Analog Signal Processing (R-ASP), which consist in processing signals in real-time using purely analog component~\cite{Caloz-2013-Analog,Rulikowski-2008-Arbitrary,Khaleghi-2011-Timereversalantenna,Ding-2013-subwavelegnth}.

In a previous work by some of the authors, time reversal sub-wavelength focusing was achieved using chirped delay lines (CDLs)~\cite{Ding-2013-subwavelegnth}. This approach removes the need for complex and large electromagnetic structures while offering at the same time superior performance. In this communication, we introduce another technique to enhance time-reversal sub-wavelength transmission. This technique is based on analog pulse shaping circuits~(PSCs), which are pulse shaping devices adding up multiple instances of the signal with different delays. Compared with ~\cite{Ding-2013-subwavelegnth}, the PSC approach offers the benefits of greater design flexibility and hence, ultimately, of higher throughput. The paper also provides an experimental validation of the technique.

The paper is organized as follows. Section~\ref{Sec:Resolution} recalls the principle of time reversal, presents the proposed time-reversal system based on PSCs, and derives the corresponding auto-correlation and cross-correlation functions for the intended and non-intended signals, as well as the PSC condition for maximal channel discrimination. Section~\ref{Sec:Discussion} defines a figure of merit for time-reversal systems, provides close-form expression for the figures of merit of PSC and CDL based systems, and demonstrates that the former is superior to the latter. Section~\ref{Sec:experiment} presents an experimental validation of the PSC based system. Finally, conclusions are given in Section~\ref{Sec:conclusion}.

\section{Channel Discrimination Enhancement \\ using Pulse Shaping Circuits (PSCs)}\label{Sec:Resolution}

\begin{figure}
\begin{center}
\includegraphics[width=1\columnwidth]{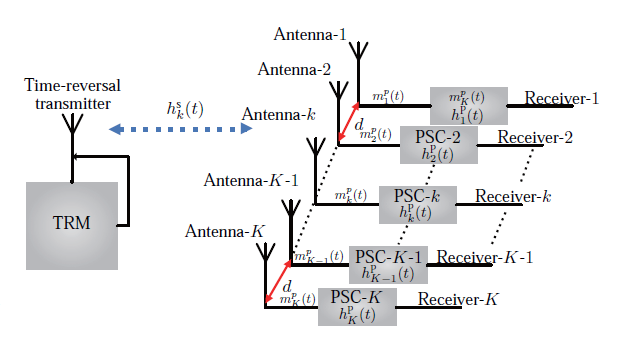}
\caption{Wireless time-reversal system using pulse shaping circuits (PSCs).}
\label{Fig:fig1}
\end{center}
\end{figure}

A wireless time-reversal system consists of a time-reversal transmitter and of a number of receivers, which may be placed at sub-wavelength distances from each other. If the distance between adjacent receivers is much less than the wavelength at the operation frequency, the system suffers from low spatial diversity, which leads to poor discrimination between the received signals. In order to solve this problem, we use here pulse shaping circuits (PSCs) with different impulse responses in each of the receivers, as shown in Fig.~\ref{Fig:fig1}. These PCSs increase the channel diversity for each receiver and hence enhance the throughput and reliability of the wireless system.

Receiver $k$ generates the channel sounding signal $s(t)$, passes it through PSC $k$, whose impulse response is $h_{k}^\text{P}(t)$ (`P' stands for PSC) for pre-processing. The pre-processed signal,
\begin{equation}
\label{eq:predistortion}
m_{k}^\text{P}(t)=s(t)\ast h_{k}^\text{P}(t),
\end{equation}
is then radiated by antenna $k$. The time-reversal transmitter receives this signal, that has traveled across the space channel, with impulse response $h_{k}^\text{s}(t)$ (`s' stands for `space'). The received signal is then
\begin{equation}
\label{eq:yk}
y_{k}(t)
=m_{k}^\text{P}(t)\ast h_{k}^\text{s}(t)
=s(t)\ast h_{k}^\text{P}(t)\ast h_{k}^\text{s}(t),
\end{equation}
where `$\ast$' denotes the convolution product.

The transmitter time-reverses this signal, and transmits the resulting signal through its antenna. The subsequently received signal by the $k^\text{th}$ receiver, to which it is destined, is
\begin{subequations}\label{eq:fp}
\begin{equation}
\label{eq:fk}
\begin{split}
f_{k}^\text{P}(t)&=s(-t)\ast h_{k}^\text{P}(-t)\ast h_{k}^\text{s}(-t)\ast h_{k}^\text{s}(t)\ast h_{k}^\text{P}(t) \\
&=s(-t)\ast \left[h_{k}^\text{P}(-t)\ast h_{k}^\text{P}(t)\right]\ast h_{k}^\text{s}(-t)\ast h_{k}^\text{s}(t)\\
&=s(-t)\ast c_{k,k}^\text{P}(t)\ast h_{k}^\text{s}(-t)\ast h_{k}^\text{s}(t),
\end{split}
\end{equation}
where
\begin{equation}\label{eq:corr}
c_{k,k}^\text{P}(t)=h_{k}^\text{P}(-t)\ast h_{k}^\text{P}(t)
\end{equation}
has been identified as the auto-correlation of $h_{k}^\text{P}(t)$, while the signal received by the $l^\text{th}$ ($l\neq k$) receiver, to which it is not destined, is
\begin{equation}
\label{eq:fl}
\begin{split}
f_{l}^\text{P}(t)&=s(-t)\ast h_{k}^\text{P}(-t)\ast h_{l}^\text{s}(-t)\ast h_{l}^\text{s}(t)\ast h_{l}^\text{P}(t)\\
&=s(-t)\ast \left[h_{k}^\text{P}(-t)\ast h_{l}^\text{P}(t)\right]\ast h_{k}^\text{s}(-t)\ast h_{l}^\text{s}(t)\\
&=s(-t)\ast c_{k,l}^\text{P}(t)\ast h_{k}^\text{s}(-t)\ast h_{l}^\text{s}(t),
\end{split}
\end{equation}
where
\begin{equation}\label{eq:corrhkl}
c_{k,l}^\text{P}(t)=h_{k}^\text{P}(-t)\ast h_{l}^\text{P}(t)
\end{equation}
\end{subequations}
has been identified as the cross-correlation of $h_{k}^\text{P}(t)$ and $h_{l}^\text{P}(t)$.

Since the the spatial channel responses $h_{k}^\text{s}(t)$ and $h_{l}^\text{s}(t)$ are highly correlated in a sub-wavelength array \cite{Ding-2013-subwavelegnth}, the following approximation can be made:

\begin{equation}\label{eq:sublamcons}
h_{k}^\text{s}(t)=h_{l}^\text{s}(t),
\end{equation}

\noindent so that the difference between $f_{k}^\text{P}(t)$ and $f_{l}^\text{P}(t)$ is entirely contained in $c_{k,k}^\text{P}(t)$ and $c_{k,l}^\text{P}(t)$, and specifically consists in the difference between $h_{k}^\text{P}(t)$ and $h_{l}^\text{P}(t)$.

The impulse responses $h_{k}^\text{P}(t)$ and $h_{l}^\text{P}(t)$ may be produced by digital signal processing (DSP) or analog signal processing (ASP). However, DSP suffers from limited speed and high power consumption due to analog-digital/digital-analog converters. Therefore, we choose here R-ASP technology~\cite{Caloz-2013-Analog}, and specifically PSCs, which will be demonstrated to be superior to CDLs~\cite{Ding-2013-subwavelegnth}.

\begin{figure}
\begin{center}
\includegraphics[width=1\columnwidth]{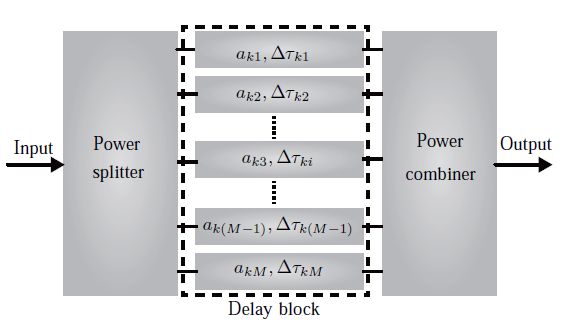}
\caption{$k^\text{th}$ PSC, with impulse response $h_k^\text{P}(t)$, in the time-reversal system of Fig.~\ref{Fig:fig1}.}
\label{Fig:fig2}
\end{center}
\end{figure}

Figure~\ref{Fig:fig2} shows the block diagram of the PSC of the $k^\text{th}$ receiver. The PSC consists of three blocks: 1)~an input 1:$M$ power splitter, which divides the input pulse into $M$ identical pulses; 2)~a delay block, consisting of $M$ delay lines which each delay the $i^\text{th}$ pulse by $\Delta\tau_{ki}$ ($i=1,2,\ldots,M$) and has a magnitude coefficient of $a_{ki}$; 3)~an output $M$:1 power combiner, which adds up the delayed pulses into a single signal. The transfer function of the $k^\text{th}$ PSC is thus
\begin{equation}
\label{eq:Hk}
H_{k}^\text{P}(\omega)
=\sum_{i=1}^{M}a_{ki}e^{-j\omega\Delta\tau_{ki}}.
\end{equation}
The transfer function of the PSC is controlled by two variable sets, the magnitude coefficient set $\{a_{ki}\}$ and the time delay $\{\Delta\tau_{ki}\}$ of each line. Here, all the magnitude coefficients are assumed to be equal,
\begin{equation}\label{eq:aki}
a_{k1}=a_{k2}=\cdots=a_{ki}=\cdots=a_{kM-1}=a_{kM}=a,
\end{equation}
and therefore the impulse response of PSC is simply
\begin{equation}\label{eq:hkpt}
h_{k}^\text{P}(t)=a\sum_{i=1}^{M}\delta(t-\Delta\tau_{ki}).
\end{equation}
Moreover, in order to facilitate the forthcoming analysis, we arrange the delay lines such that
\begin{equation}\label{eq:tau}
\Delta\tau_{k1}<\Delta\tau_{k2}<\cdots<\Delta\tau_{ki}<\Delta\tau_{kM-1}<\Delta\tau_{kM}.
\end{equation}
As previously mentioned, the level of discrimination between $f_{k}^\text{P}(t)$ and $f_{l}^\text{P}(t)$ depends on the degree of dissimilarity between $c_{k,k}^\text{P}(t)$ and $c_{k,l}^\text{P}(t)$.
Substituting~\eqref{eq:hkpt} into~\eqref{eq:fp} and using the even property of the Dirac delta function yields
\begin{subequations}\label{eq:exc}
\begin{equation}\label{eq:exckk}
\begin{split}
c_{k,k}^\text{P}(t)&=a^{2}\sum_{i=1}^M\sum_{j=1}^{M}\delta(t+\Delta\tau_{ki})\ast\delta(t-\Delta\tau_{kj})\\
&=a^{2}\sum_{i=1}^M\sum_{j=1}^{M}\delta(t-\Delta\tau_{kj}+\Delta\tau_{ki})\\
&=Ma^{2}\delta(t)+a^{2}\sum_{i=1}^M\sum_{j=1,j\neq i}^{M}\delta(t-\Delta\tau_{kj}+\Delta\tau_{ki})
\end{split}
\end{equation}
and
\begin{equation}\label{eq:exckl}
\begin{split}
c_{k,l}^\text{P}(t)&=a^{2}\sum_{i=1}^M\sum_{j=1}^{M}\delta(t+\Delta\tau_{ki})\ast\delta(t-\Delta\tau_{lj})\\
&=a^{2}\sum_{i=1}^M\sum_{j=1}^{M}\delta(t-\Delta\tau_{lj}+\Delta\tau_{ki})\\
&=a^{2}\sum_{i=1}^{M}\delta(t-\Delta\tau_{li}+\Delta\tau_{ki})\\
&\qquad+a^2\sum_{i=1}^M\sum_{j=1,j\neq i}^{M}\delta(t-\Delta\tau_{lj}+\Delta\tau_{ki}).
\end{split}
\end{equation}
\end{subequations}
By definition of the auto-correlation, $c_{k,k}^\text{P}(t)$ reaches its peak (maximum) at $t=0$. Consider the cross-correlation, $c_{k,l}^\text{P}(t)$. If
\begin{equation}\label{eq:taun}
\Delta\tau_{l1}-\Delta\tau_{k1}=\cdots=\Delta\tau_{li}-\Delta\tau_{ki}=\cdots=\Delta\tau_{lM}-\Delta\tau_{kM}=C_{kl},
\end{equation}
where $C_{kl}$ is independent of $i,j$, $\forall k,l$, then
\begin{equation}\label{eq:cklnc}
c_{k,l}(t)=Ma^{2}\delta(t-C_{kl})+a^{2}\sum_{i=1}^M\sum_{j=1,j\neq i}^{M}\delta(t-\Delta\tau_{kj}+\Delta\tau_{ki}).
\end{equation}
In this case, comparison with~\eqref{eq:exckk} indicates that the peak of $c_{k,l}(t)$ will be as high as the peak of $c_{k,k}(t)$\footnote{Only the first term in~\eqref{eq:cklnc} and ~\eqref{eq:exckk} contribute to the peak of the function since in the second terms the Dirac delta function terms lead to maxima that are distributed in time.}, which means that $h_{k}^\text{P}(t)$ and $h_{l}^\text{P}(t)$ are fully correlated, and that there is therefore no channel discrimination. If, in contrast,
\begin{equation}\label{eq:taun}
\Delta\tau_{l1}-\Delta\tau_{k1}\neq\cdots\neq\Delta\tau_{li}-\Delta\tau_{ki}\neq\cdots\neq\Delta\tau_{lM}-\Delta\tau_{kM},
\end{equation}
the maxima in the first term of~\eqref{eq:exckl} are completely spread out in time (all the terms contribute different times), then this first term is much smaller than the first term in~\eqref{eq:exckk} and therefore $h_{k}^\text{P}(t)$ and $h_{l}^\text{P}(t)$ are minimally correlated, which leads to maximal channel discrimination.

\section{Comparison Between PSC and CDL based Time-reversal Systems}\label{Sec:Discussion}

\subsection{Comparative Figure of Merit (FOM)}

For comparing the PSC and DCL time-reversal approaches, one must establish a proper comparison basis. This basis will be provided here by the derivation of a transmission figure of merit.

The PSC and DCL systems will be both assumed to use as the information carrier signal, $s(t)$, the Gaussian pulse
\begin{equation}
\label{eq:Gaussian}
g(t)=\text{exp}\left(-\frac{t^2}{2T_{0}^{2}}\right)\text{cos}(\omega_0t),
\end{equation}
where $T_0$ is the full time width at half maximum and $\omega_0$ is the modulated pulse frequency of the pulse. Typical modulation schemes based on this pulse could be on-off keying (OOK) or pulse position modulation (PPM).

According to \eqref{eq:fp}, the signals detected by the $k^\text{th}$ and $l^\text{th}$ receivers generally read, using $s(t)=g(t)$ and the even property of $g(t)$,
\begin{subequations}\label{eq:ftp}
\begin{equation}
\label{eq:ftk}
\begin{split}
f_{k}^\text{T}(t)
&=\left[g(t)\ast h_{k}^\text{T}(-t)\ast h_{k}^\text{T}(t)\right]\ast h_{k}^\text{s}(-t)\ast h_{k}^\text{s}(t)\\
&=u_{k,k}^{T}\ast h_{k}^\text{s}(-t)\ast h_{k}^\text{s}(t),
\end{split}
\end{equation}
with
\begin{equation}\label{eq:corrfkk}
u_{k,k}^\text{T}(t)=g(t)\ast h_{k}^\text{T}(-t)\ast h_{k}^\text{T}(t),
\end{equation}
and
\begin{equation}
\label{eq:ftl}
\begin{split}
f_{l}^\text{T}(t)
&=\left[g(t)\ast h_{k}^\text{T}(-t)\ast h_{l}^\text{T}(t)\right]\ast h_{k}^\text{s}(-t)\ast h_{l}^\text{s}(t)\\
&=u_{k,l}^{T}\ast h_{k}^\text{s}(-t)\ast h_{l}^\text{s}(t),
\end{split}
\end{equation}
with
\begin{equation}\label{eq:corrfkl}
u_{k,l}^\text{T}(t)=g(t)\ast h_{k}^\text{T}(-t)\ast h_{l}^\text{T}(t),
\end{equation}
where the superscript `T' stands for either PSC or CDL.
\end{subequations}

We may then define the transmission figure of merit, at the transmission maximum, as
\begin{equation}\label{Eq:FOM}
F^\text{T}=\frac{\text{max}\{|u_{k,k}^\text{T}(t)|^2\}}{\text{max}\{|u_{k,l}^\text{T}(t)|^2\}}.
\end{equation}

\noindent This quantity measures the level of power discrimination between the destined and non-destined signals.

\subsection{PSC FOM}

Under the minimum correlation condition~\eqref{eq:taun}, Eqs.~\eqref{eq:corrfkk} and \eqref{eq:corrfkl} become, for the PSC case,
\begin{subequations}\label{eq:PCGA}
\begin{equation}\label{eq:PGA}
u_{k,k}^\text{P}(t)=Ma^2g(t)+a^{2}\sum_{i=1}^{M}\sum_{j=1,i\neq j}^{M}g(t-\Delta\tau_{kj}+\Delta\tau_{ki})
\end{equation}
and
\begin{equation}\label{CGA}
u_{k,l}^\text{P}(t)=a^{2}\sum_{i=1}^{M}\sum_{j=1}^{M}g(t-\Delta\tau_{lj}+\Delta\tau_{ki}).
\end{equation}
\end{subequations}

Assuming $T_0<\Delta\tau_i$ (no overlap between adjacent bit pulses), the PSC figure of merit is found, by inserting~\eqref{eq:PCGA} into~\eqref{Eq:FOM}, as
\begin{equation}
F^\text{P} = \frac{\text{max}\{|u_{k,k}^\text{P}(t)|^2\}}{\text{max}\{|u_{k,l}^\text{P}(t)|^2\}} = \left(\frac{Ma^2}{a^2}\right)^2 = M^2.
\label{Eq:DP_PSC}
\end{equation}

\subsection{CDL FOM}

In the CDL case, maximal channel discrimination is achieved when the channels have opposite opposite group delay slopes~\cite{Ding-2013-subwavelegnth}. Since the CDL slopes must be equal in magnitude, for proper signal retrieval, the CDL system is inherently limited to two channels.

According to \cite{Agrawal-Optical-Book, Babak-2011-Enhance}, Eqs.~\eqref{eq:corrfkk} and \eqref{eq:corrfkl} become, for the CDL case,
\begin{subequations}\label{eq:autocc}
\begin{equation}\label{eq:autocukk}
u_{k,k}^\text{C}(t)=\text{exp}\left(-\frac{t^2}{2T_{0}^{2}}\right)\text{cos}(\omega_0t)
\end{equation}
and
\begin{equation}\label{eq:autocukl}
u_{k,l}^\text{C}(t)=\frac{T_{0}}{\sqrt{T_{0}^{2}-2j\sigma}}\text{exp}\left\{-\frac{t^2}{2(T_{0}^{2}-2j\sigma)}\right\}\text{cos}(\omega_0t),
\end{equation}
\end{subequations}
where $\sigma$ is the CDL group delay slope, and `C' stands for `CDL'.

The corresponding figure of merit is
\begin{equation}
F^\text{C} = \frac{\text{max}\{|u_{k,k}^\text{C}(t)|^2\}}{\text{max}\{|u_{k,l}^\text{C}(t)|^2\}} = \frac{\sqrt{T_0^4 + 4\sigma^2}}{T_0^2}.
\label{Eq:DP_CDL}
\end{equation}

\subsection{Comparison}

\begin{figure}
\begin{center}
\includegraphics[width=0.8\columnwidth]{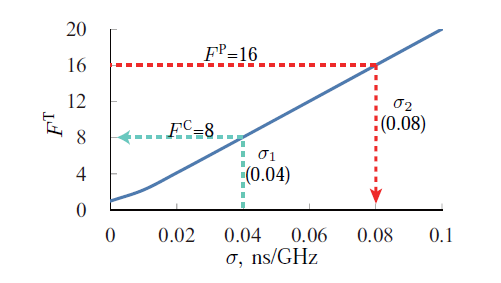}
\caption{Compared figures of merit for the PSC ($M=4$ delay lines) and CDL time-reversal systems calculated by~\eqref{Eq:DP_PSC} and~\eqref{Eq:DP_CDL} for $K=2$~channels.} \label{Fig:PSC_Bragg_R}
\end{center}
\end{figure}

\begin{figure}
\begin{center}
\includegraphics[width=\columnwidth]{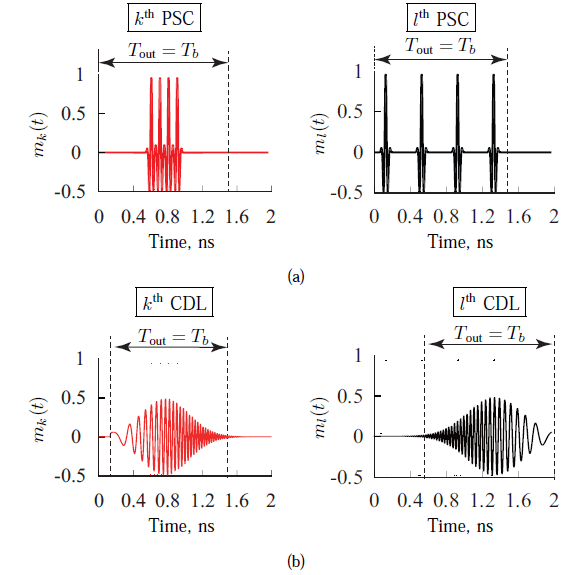}
\caption{Normalized pre-processed signals $m(t)$. a)~PSC case. b)~CDL case.} \label{Fig:OP_PSC_BG}
\end{center}
\end{figure}

\begin{figure}
\begin{center}
\includegraphics[width=0.9\columnwidth]{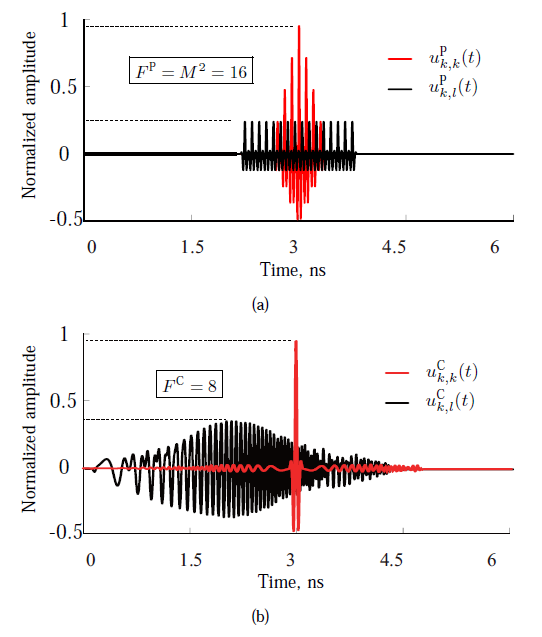}
\caption{Auto-correlations and cross-correlations using the signals of Fig.~\ref{Fig:OP_PSC_BG}. a)~PSC case. b)~CDL case with $\sigma=\sigma_1=0.04$ ns/GHz. The pre-processed pulse duration is $T_m=1.5$~ns, corresponding to equal throughput, for the two signals.} \label{Fig:PSC_CDL_SameTb}
\end{center}
\end{figure}

Figure~\ref{Fig:PSC_Bragg_R} provides a graphical comparison between the PSC and CDL systems, for typical microwave values of $\sigma$~\cite{Schwartz-2010-Chirped}. This figure assumes equal durations for the seed sounding pulses, $T_0=0.1$~ns, and for the pre-processed sounding pulses, $T_m=1.5$~ns, for the two systems, as shown in Fig.~\ref{Fig:OP_PSC_BG}, meaning that the time-expansion is of $T_m-T_0=1.4$~ns for both cases. This assumption ensures equal throughput for the two systems. The $1.4$~ns expansion corresponds to $\sigma_1=0.04$~ns in the CDL system, indicated in Figure~\ref{Fig:PSC_Bragg_R} and shown in the same figure to correspond to a figure of merit of $F^\text{C}=8$. Under the aforementioned identical pulse durations, the PSC based system with $M=4$ provides a figure of merit of $F^\text{C}=16$, and is therefore $3$~dB superior to the CDL based system.
In addition, Fig.~\ref{Fig:PSC_CDL_SameTb} shows that the transmitted pulse duration of the CDL based system is about three times larger than that of the PSC based system, since the pulse duration must include both the auto-correlation cross-correlation signals to avoid excessive cross-correlation energy that may result from overlapping adjacent bits, indicating that the throughput of the latter is three times larger.
One may boost the CDL figure of merit to $16$ (level of the PSC ) by increasing $\sigma$ to $\sigma_2=0.08$. However, the resulting transmitted pulse duration is then further increased, to 6~ns, which is four times the duration of the PSC transmitted pulse duration, and corresponds therefore to a four times smaller throughput.
The PSC based system is thus superior to the CDL based system in microwave technology. And in case higher-resolution CDL system could be engineered, the PSC could maintain its superiority by simply using more delay lines.
\begin{figure}
\includegraphics[width=1\columnwidth]{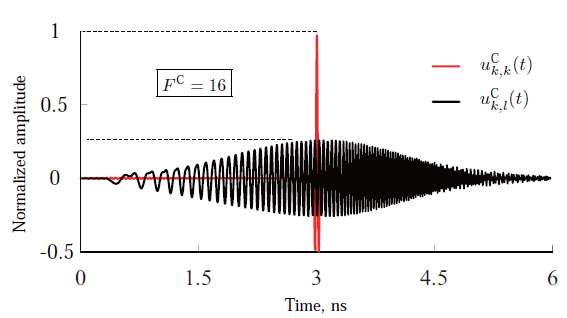}
\caption{CDL output pulses and figure of merit for $\sigma=\sigma_2=0.08$~ns.}
\label{Fig:PSC_CDL_SameR}
\end{figure}

\section{Experiment Demonstration}\label{Sec:experiment}

Figure~7(a) shows the schematic experimental setup, corresponding to~Fig.~\ref{Fig:fig1} for $K=2$ channels and $M=4$ PSC delay lines, as in the numerical example of Sec.~\ref{Sec:Discussion}. The experiments are carried out in an indoor environment. A modulated Gaussian pulse with $5$~GHz central frequency and $3$~GHz bandwidth is selected as the channel sounding pulse. The time-reversal transmitter uses an arbitrary waveform generator (Tektronix AWG7122B) to emulate the time-reversed sounding pulses and a $3.1-10.6$~GHz UWB antenna for transmission. The two receivers receive the signal using UWB antennas, shown in Fig.~7(b) (similar to those of the transmitter), incorporate the PSCs shown in Fig.~7(c) for pulse shaping, and are connected to a digital serial analyzer (Tektronix DSA72004B) for data analysis. The distance between two adjacent receiving antennas is $10$~mm, which is far less than the wavelength of the operation: $f=6.5$ to $3.5$~GHz $\rightarrow$ $\lambda_0=46$ to $86$~mm  $\rightarrow$ $d=\lambda_0/46$ to $\lambda_0/86$.
\begin{figure}
\begin{center}
\includegraphics[width=0.8\columnwidth]{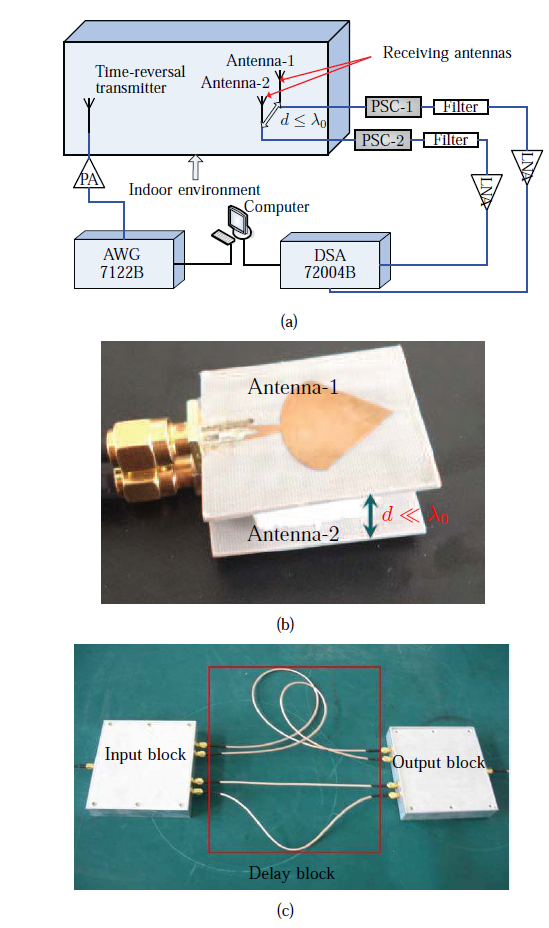}
\caption{Experiment. a) Schematic of the setup. b) Receiving antennas. c) One of the two pulse shaping circuits (PSCs).}\label{Fig:setup}
\end{center}
\end{figure}
The PSCs [Fig.~\ref{fig:loop3}] are built as follows. An input 1:4 power splitter (DC to $8$~GHz operation frequency range) splits the input pulse into four identical pulses. The same device is used as the output 4:1 power combiner to sum up the delayed pulses. The delay block consists of $4$ delay lines of different lengths selected so as to maximize channel discrimination according to~\eqref{eq:taun}. The delay times of the delay lines satisfy the conditions $\Delta\tau_{k,i+1}-\Delta\tau_{k,i}=0.9$ ns and $\Delta\tau_{l,i+1}-\Delta\tau_{l,i}=0.3$ ns.
The same equivalent channel sounding method, based on channel reciprocity, as in~\cite{Ge-2011-subwavelegnth,Ding-2013-subwavelegnth}, is used in the calibration phase, where the sounding signals are actually generated by the transmitter and measured by the receivers, for simplicity. The experimental procedure is as follows:
\begin{enumerate}
  \item generate the channel sounding pulse, using the arbitrary wave generator, and transmit it with the time-reversal transmitter;
  \item record the signals received by the receiving antennas using the digital serial analyzer, flip them and numerically modulate them using a computer;
  \item transmit the modulated time-reversed signals from the time-reversal transmitter and record the signals received by both the target antenna and the non-target antenna.
\end{enumerate}
The experimental results are shown in Fig.~\ref{Fig:Case2}. Thanks to the PSCs, the waveform of the signal received by the target receiver is much higher and essentially identical to the transmitted one (not shown) while the signal received by the non-target receiver has a totally different waveform and is spread out in time with much lower temporal power density.
\begin{figure}
\begin{center}
\includegraphics[width=1\columnwidth]{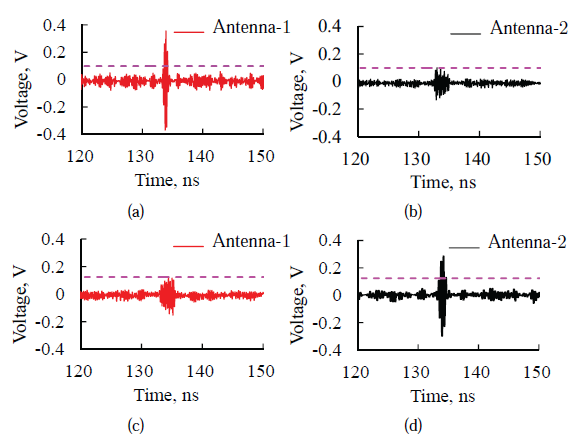}
\caption{Signals received by the antennas (experiment). a) and b) denote the case when antenna-$1$ is set to be the target antenna. c) and d) denote the case when antenna-$2$ is set to be the target} \label{Fig:Case2}
\end{center}
\end{figure}
\begin{figure}
\begin{center}
\includegraphics[width=1\columnwidth]{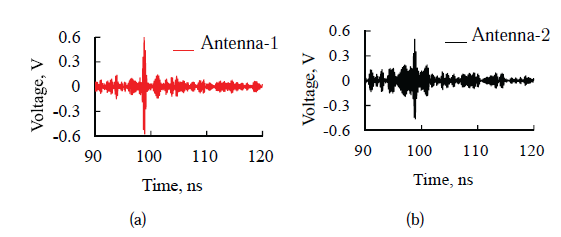}
\caption{Same experiment as in Fig.7(a) but without pulse shaping circuits. a) Signal received by Antenna-1. b) Signal received by Antenna-2.} \label{Fig:compare}
\end{center}
\end{figure}

For comparison, the same experiment is carried out without PSCs. The results are shown in Fig.~\ref{Fig:compare}. In this case, as expected, the waveform of the signal received by the target receiver has lost its target features and is comparable to that received by the non-target receiver.

\section{Conclusions}\label{Sec:conclusion}

A novel approach to enhance time-reversal subwavelength transmission based on R-ASP PSCs has been proposed, theoretically derived and experimentally validated. The PSC approach has been demonstrated to be superior to the CDL approach previously reported by some of the authors.

Due to their inherent analog nature and high performance characteristics, PSC based time-reversal sub-wavelength transmission system may find wide applications in compact MIMO and high resolution imaging.

\bibliographystyle{IEEEtran}
\bibliography{ReferenceList}

\end{document}